\documentclass{epl}
\usepackage[latin1]{inputenc}
\title{Dispersive charge transport along the surface of an insulating layer observed by Electrostatic Force Microscopy}
\author{J. Lambert \inst{1}  \and G. de Loubens \inst{2} \and T. Mélin \inst{2} \and C. Guthmann \inst{1} \and M. Saint-Jean \inst{1} }
\institute{
  \inst{1} Groupe de Physique des Solides - 2, place Jussieu, Tour 23, 75251 Paris Cedex 05, France\\
  \inst{2} Institut d'Electronique de Microélectronique et de Nanotechnologie - CNRS UMR 8520 - Avenue Poincaré, BP 69, 59652 Villeneuve d'Ascq Cedex, France\\
} 
\begin{document}

\maketitle

\begin{abstract}
We report the observation in the direct space of the transport of a few thousand charges submitted to a
tunable electric field along the surface of a silicon oxide layer. Charges are both deposited and observed
using the same Electrostatic Force Microscope. During the time range accessible to our measurements (i.e.
$t=1\sim1000\un{s}$), the transport of electrons is mediated by traps in the oxide. We measure the mobility
of electrons in the "surface" states of the silicon oxide layer and show the dispersive nature of their motion. It is also demonstrated that the saturation of deep oxide traps strongly enhance the transport of electrons under lateral electric field.
\end{abstract}
\section{Introduction}
Silicon dioxide is nowadays widely used as an insulating material in microelectronic devices: Its homogeneity, and thus its lack of impurities, together with its wide band gap, make it a very good insulating material. However one must notice that electronic devices based on silicon dioxide are very sensitive to static discharges due to extrinsic charges. Therefore one might ask a first question of technological importance :  what happens to a few charges on the surface of such a good insulating material ? Will the charges move across the insulating layer towards the conductive underlying substrate or will they move along the surface towards a polarized electrode under the influence of the induced electric field ? We show here, using an Electrostatic Force Microscope, that both kinds of behaviors may be observed and that such a charge distribution is mobile under the influence of an electric field; this latter being low enough to be induced in a normal electronic device such as a MOSFET. \\

From the fundamental point of view, observing, in the direct space, the motion of a charge packet under a constant electric
field in an amorphous insulating layer, be it silicon dioxide or another amorphous insulating material like
any high-k dielectric, represents a fundamental topic. Many experiments performed to study the
motion of charges in insulating layers involve the measurement of a transient current across the layer, this
current is created by the motion of excited electrons or holes. The current integrates the contributions of
the whole charge distribution over the thickness of the layer \cite{zallen,batra,wintle,kanazawa}.
In the case of classical diffusive transport, the measurement of the mobility is sufficient to characterize entirely the motion of charges, and is generally measured macroscopically from the charge "time of flight" across the layer \cite{zallen, scher}.
Actually, for a wide class of materials
one observes a continuous decay of the current, indicating a spreading of the individual
"time-of-flight", which reflects the fact that the motion of the charges in insulating materials involves
successive stochastic trappings and detrappings of the charges in localized states.
\\
It may occur that the asymptotic behavior of the current at small and large times may be fitted by power-law
functions. This kind of behavior is characteristic of dispersive transport. A first interpretation for this
kind of transport phenomenon was given by Scher and Montroll\cite{scher} who demonstrated that the
whole transport process was controlled by long trapping times.
As a consequence, the spreading of the charge packet is proportional to the mean charge displacement, and
follows a time-linear law. Let us underline, after Arkhipov \textit{et al.} that dispersive transport occurs
in the frame of the transient motion of charges, when a thermalisation of mobile charges still did not occur
\cite{arkhipov2001,aetr1,aetr2,aetr3}. Therefore the second question we address in this
letter is whether it is possible or not to observe the spreading of a charge distribution in the direct
space, and thus to determine without ambiguity the transport regime involved in the motion of the charges.

Our strategy was to perform charge deposits and study them using the same EFM under the influence of an electric field applied along the surface of the insulating layer.


\section{Experimental apparatus}
We chose to perform the following experiments using thermal silicon dioxide layers as insulating materials, first and foremost because of their technological interest; but also because their integration in micro-electronic devices has induced the mastering of their homogeneity together with the possibility to incorporate electrodes and contacts in them at any stage of their making. This latter point was crucial in our experiments since the 
electrodes had to be buried in the insulator in order to minimize
their electric influence on the EFM tip and ensure a maximum transverse 
field within the insulating layer.
Samples were prepared from a 400 nm thick thermal silicon dioxide layer 
grown on a
p-type doped silicon wafer. Interdigitated electrode arrays 
with
electrode spacing of $1\un{\mu m}$, $3\un{\mu m}$ and $10\un{\mu m}$
were achieved.
EFM charge injection experiments were found similar
for all patterned samples as well as for un-patterned silicon dioxide 
layers.
On patterned samples, the size of the deposits on the surface, about 500 nm wide \cite{saintjean3}, have brought us to use mainly $10\un{\mu m}$ distant electrodes. In this configuration, the field induced at the surface of the silicon dioxide layer is approximately $10^{4} \un{V.m^{-1}}$, a value that is weak compared to the dimensions of the electrodes and to the lateral bias $V^{+} - V^{-}$. This is due to the fact that the conducting doped silicon substrate (counter-electrode) screens the field induced by the embedded electrodes.
Hereafter this electric field will be designated as a "transverse" electric field in contrast with the field induced by the tip in the insulating layer.\\

In order to deposit and to measure charges of both signs on the $\chem{SiO_2}$ surface with the same instrument, we developed an home made modified electrostatic force microscope (EFM). A complete description of the different functioning modes of  this instrument can be found in \cite{nordin}. Let us here remind the main characteristics of this set-up. The localized charges are deposited by contact electrification : the tip is brought in contact with the $\chem{SiO_2}$ layer and a deposit voltage is applied between the metallic tip and the conductive substrate of the insulating layer during a short time. One can choose the quantity of deposited charges  and their spatial distribution by tuning the deposit voltage and the deposit time \cite{saintjean2}. Once the deposit is performed, the tip is withdrawn from the surface. In order to determine the quantity of deposited charges and their evolution, we measure the variations of the force applied on the tip when it scans the charged  surface. \\

\begin{figure}
\onefigure[width=8cm]{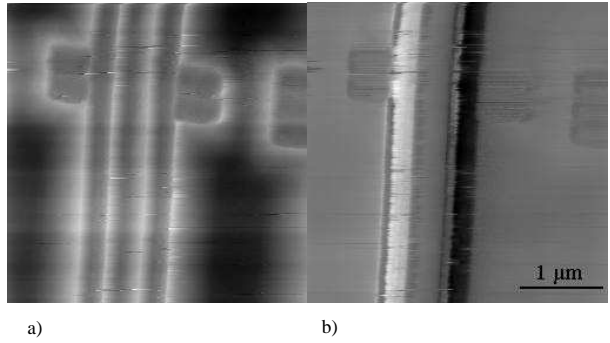} \caption{EFM images of biased electrodes:
a) Topographical signal,
b) Electrical signal.
The two electrodes are put respectively at $+2.85 \un{V}$ and $-2.85\un{V}$. 
} \label{f.2}
\end{figure}

In order to select the electrostatic force as the dominant force contribution, this distance has to be greater than $20\un{ nm}$\cite{saintjean1}, and in this case the measured force is weak (a few tenth of pN), therefore one usually uses a resonant method.
In this operating mode, the tip oscillates at an amplitude of a few angströms, the cantilever being both excited by a bimorph near its resonance frequency and by an electrical force resulting from an alternative potential applied between the tip and the conductive substrate of the insulating layer at a frequency $\Omega\approx 10 \un{kHz} $. The force exerted on the tip is then a sum of three components oscillating at frequencies $\omega_{cantilever}$ , $\Omega$ and $2\Omega$ respectively:

$F_{0}=\frac{\partial_{z}C(z)}{2}[(V_{0}+V_{\sigma})^{2}+\frac{V_{\Omega}^{2}}{2}]$, $F_{\Omega}=\partial_{z}C(z) V_{\Omega}(V_{0}+V_{\sigma})$ and $F_{2\Omega}=\frac{\partial_{z}C(z) V_{\Omega}^{2}}{4}$;
where $V_{\sigma}$ contains the contributions to the potential on the tip apex due to the embedded charges. By using a feedback loop keeping the value of $\partial F_{0}/\partial z$ constant during the scan of the tip above the deposited charges, one may measure simultaneously the variations of z, and of $F_{\Omega}$ which is proportional to $V_{\sigma}$ since the variations of the applied forces on the tip induced small modifications of its oscillation amplitude which may be measured above each point of the surface by an interferometer. Hence the force profiles for each frequency can be mapped. Once these images are recorded, the main question is to extract the corresponding charge distribution. This point requires a great care \cite{lambert2} and may be obtained by setting $V_{0}+V_{\sigma}< V_{\Omega}$.
As an illustration of this latter point, Fig.~\ref{f.2} shows the topographical and electrical signals
recorded above polarized electrodes buried in a silicon dioxide layer. 

\section{Experimental results and interpretation}

\begin{figure}[ht]
    \begin{center}
\includegraphics[width=7cm]{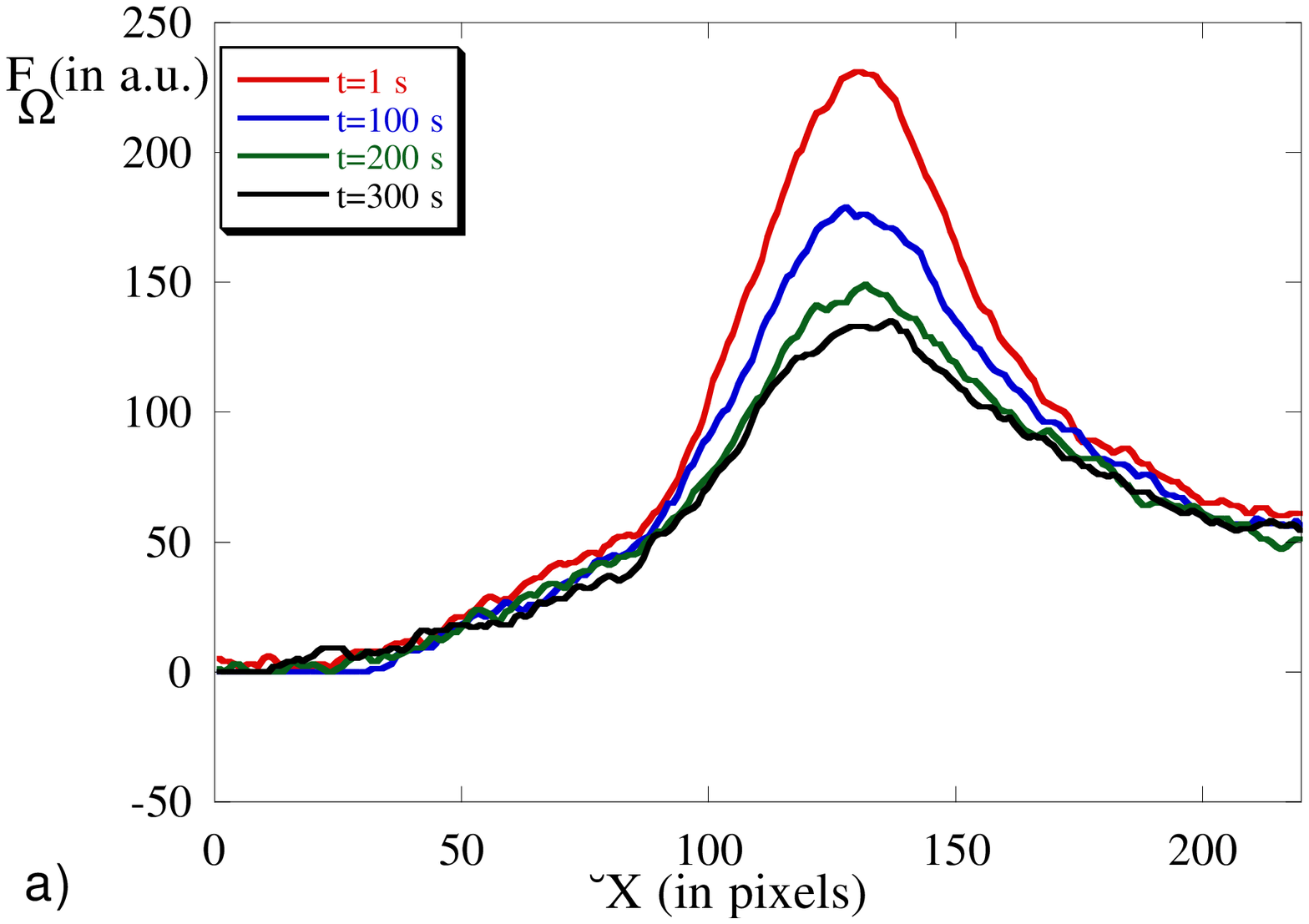}
\includegraphics[width=7cm]{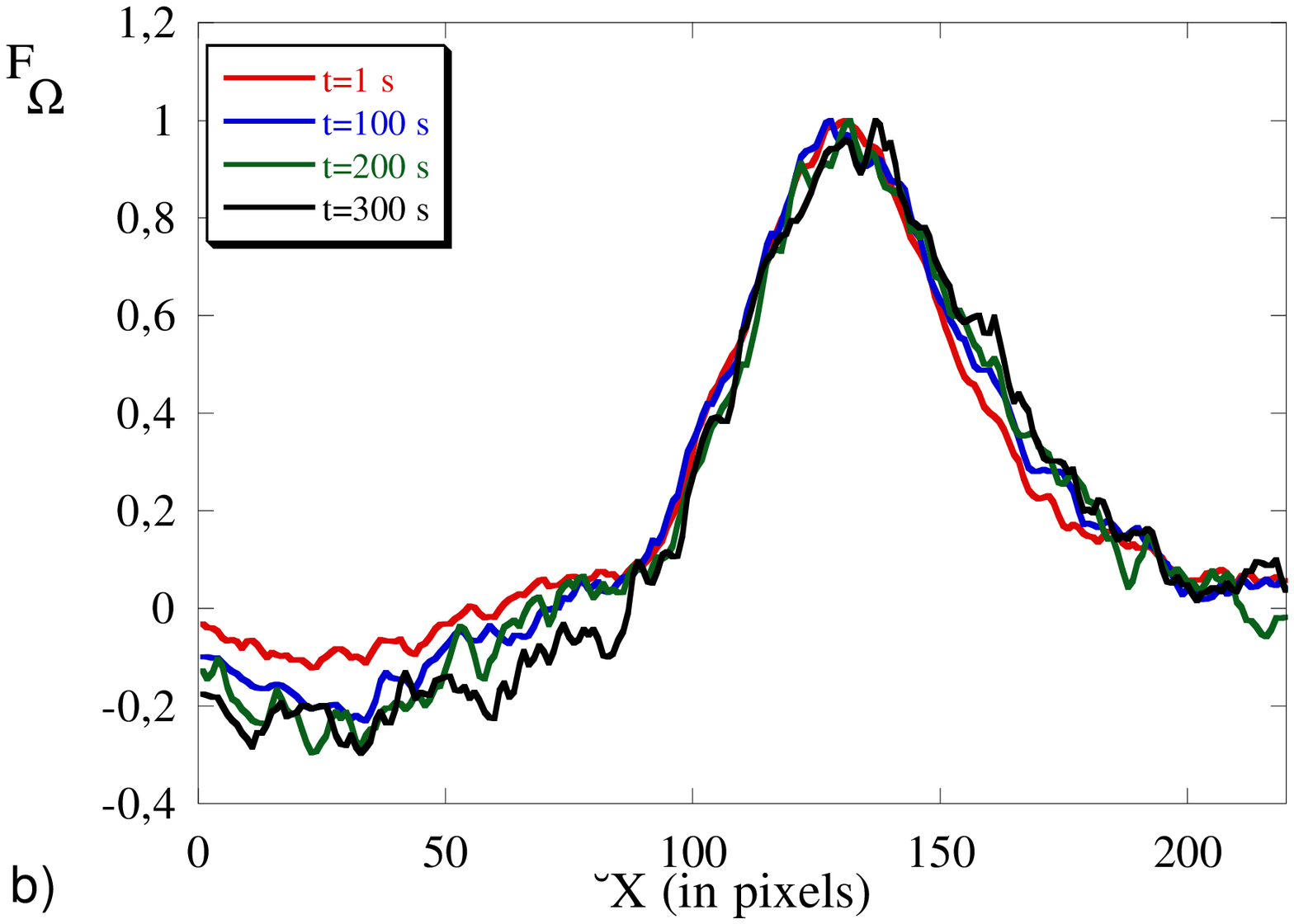}

\caption{(a): Electrical profiles recorded above a charge distribution. The deposit was performed at $V_{d}=-40\un{V}$ for times $1\un{s}$, $100\un{s}$, $200\un{s}$ and $300\un{s}$ after the deposit. The total width is $1\un{\mu m}$. (b): Same profiles after normalization.  }\label{f.3}
\end{center}
\end{figure}
Let us first describe the typical relaxation of charges upon which no external electric field is exerted.  A positive
charge packet is deposited by contact electrification ($V_{d}=-40\un{V}$) on a thick \chem{SiO_{2}} layer. The evolution of the profile of this charge
distribution is recorded, starting $1\un{s}$  after the charge deposit, and is shown on Fig~\ref{f.3}. One may first notice that the width of the charge distribution is considerably larger than the radius of
the apex of the EFM tip. In this situation, one may consider that the force exerted on the tip is strictly
proportional to the charge density under the tip integrated over the whole thickness of the layer (see e.g.
\cite{saintjean3,lambert2}). The total charge quantity deposited is evaluated at a few thousand charges. The second important experimental fact is
the absence of spreading of the charge distribution during its relaxation as can be seen on figure 3b, where
the charge profile measured at different steps of its relaxation has been normalized: hence it is clear that
the dominant mechanism for charge relaxation occurs across the layer. An interpretation of this phenomenon
will be given afterwards and in \cite{enfoncement}.

\begin{figure}[h]
\begin{center}
\includegraphics[width=7cm]{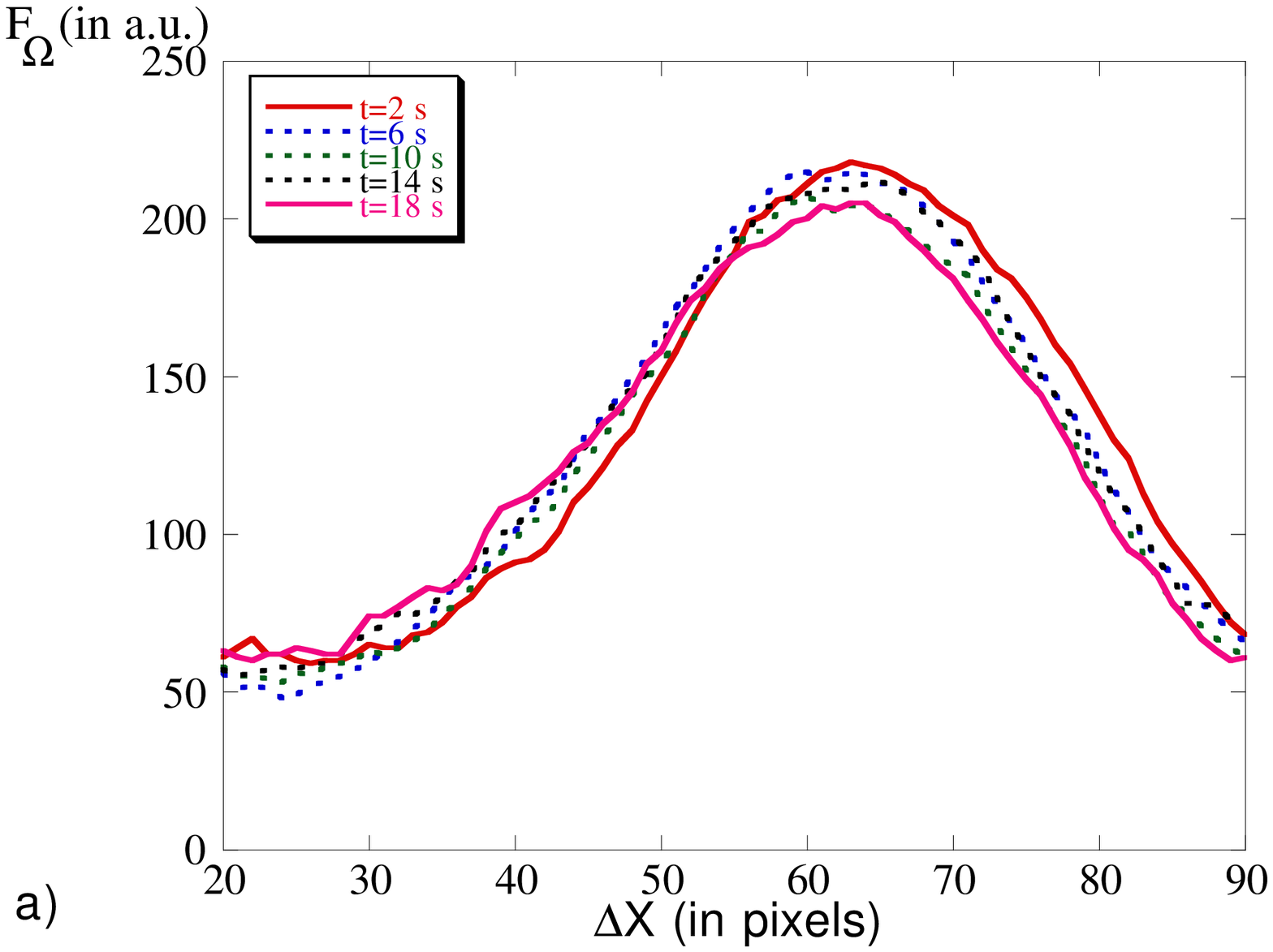}
\includegraphics[width=7cm]{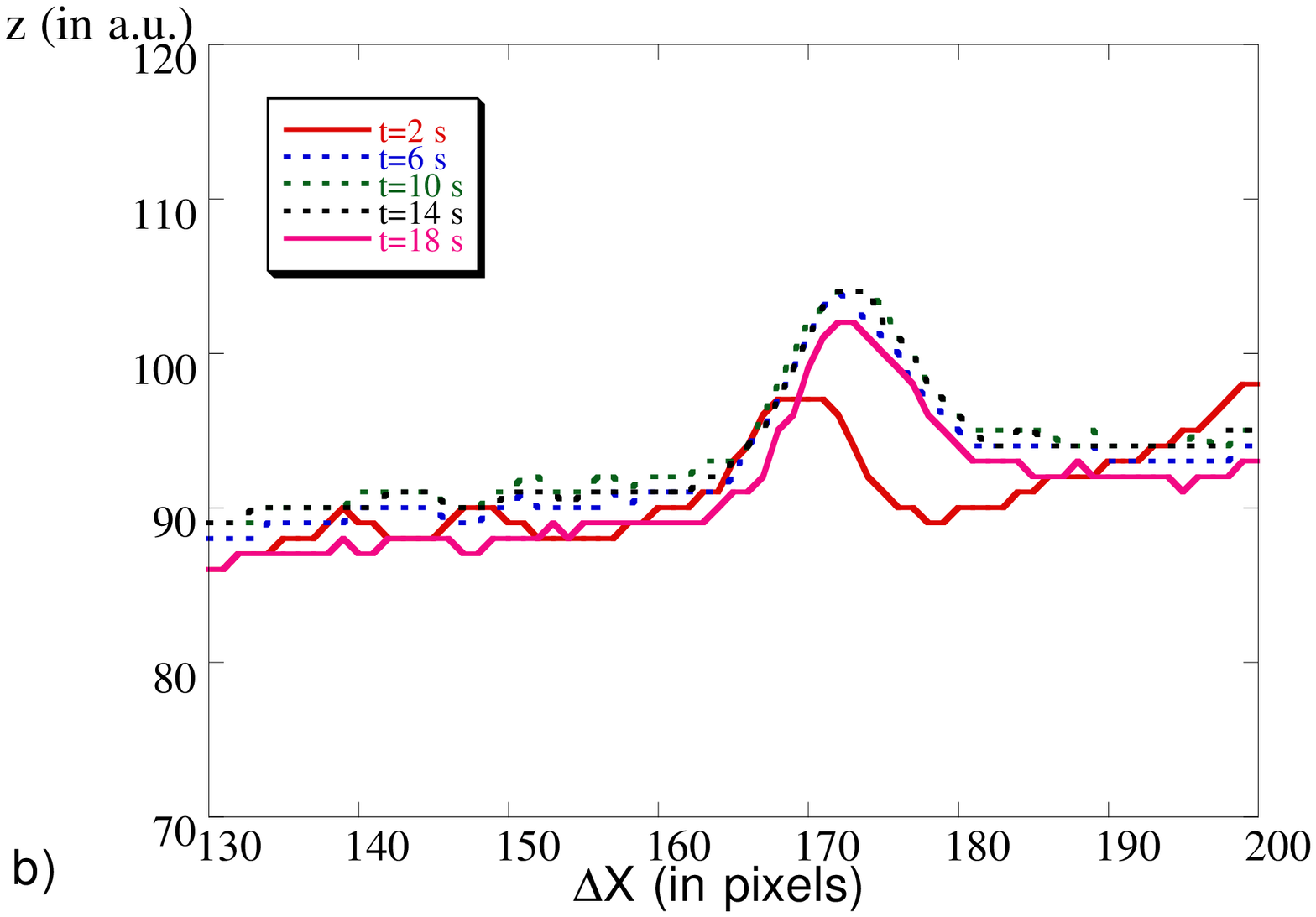}
\caption{Temporal evolution of the electrical (a) and topographical (b) profiles measured by EFM over a $1.5\un{\mu m}$ segment of the \chem{SiO_2} surface. The deposit was performed using $V_{d} = +50 V$ and $t_{d} = 200\, ms$. The electrodes are polarized at $+6,5 \un{V}$ and $-6,5 \un{V}$ respectively.} \label{f.4}
\end{center}
\end{figure}
\begin{figure}
    \begin{center}
\includegraphics[width=4cm]{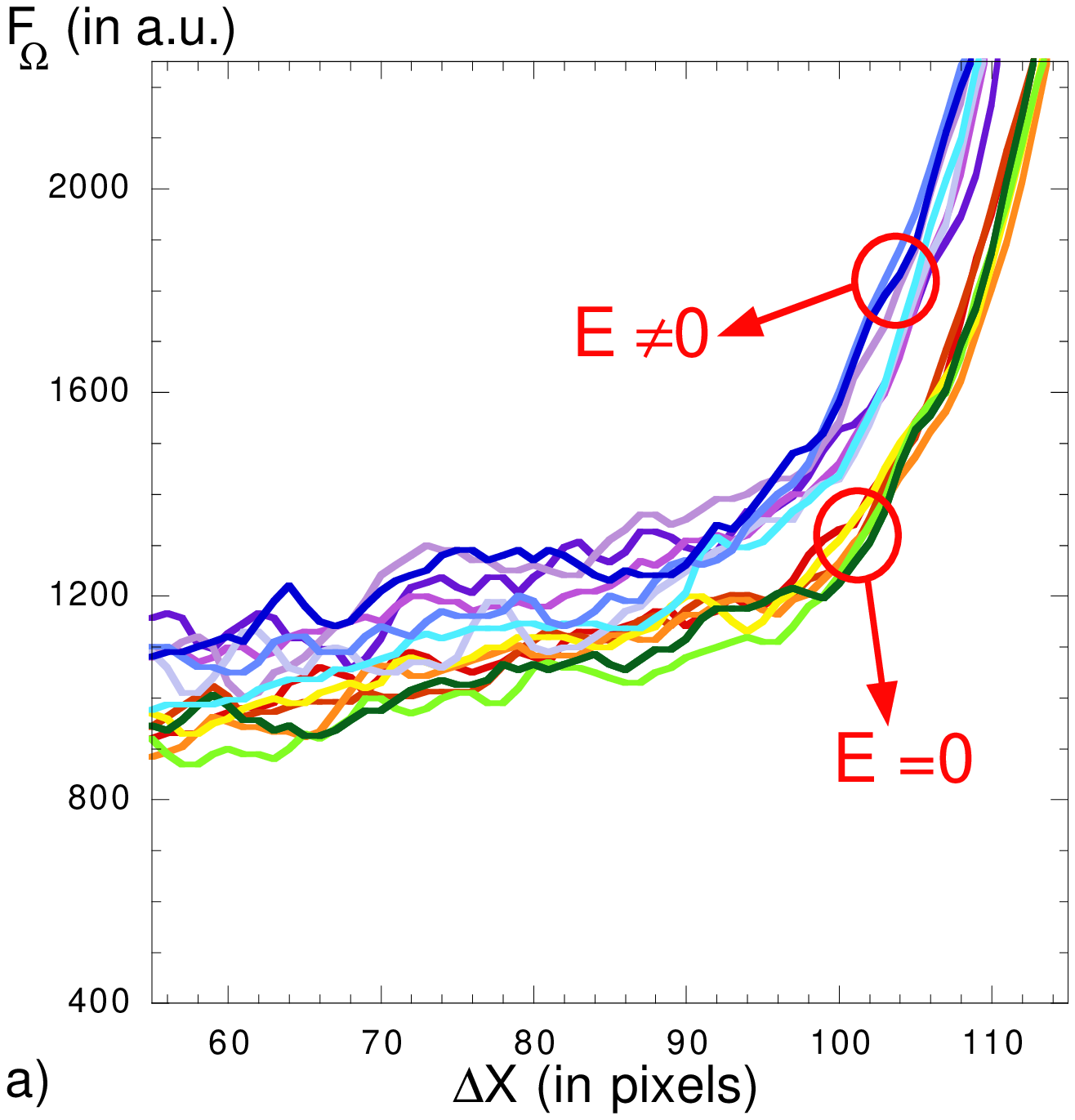}
\includegraphics[width=3.6cm]{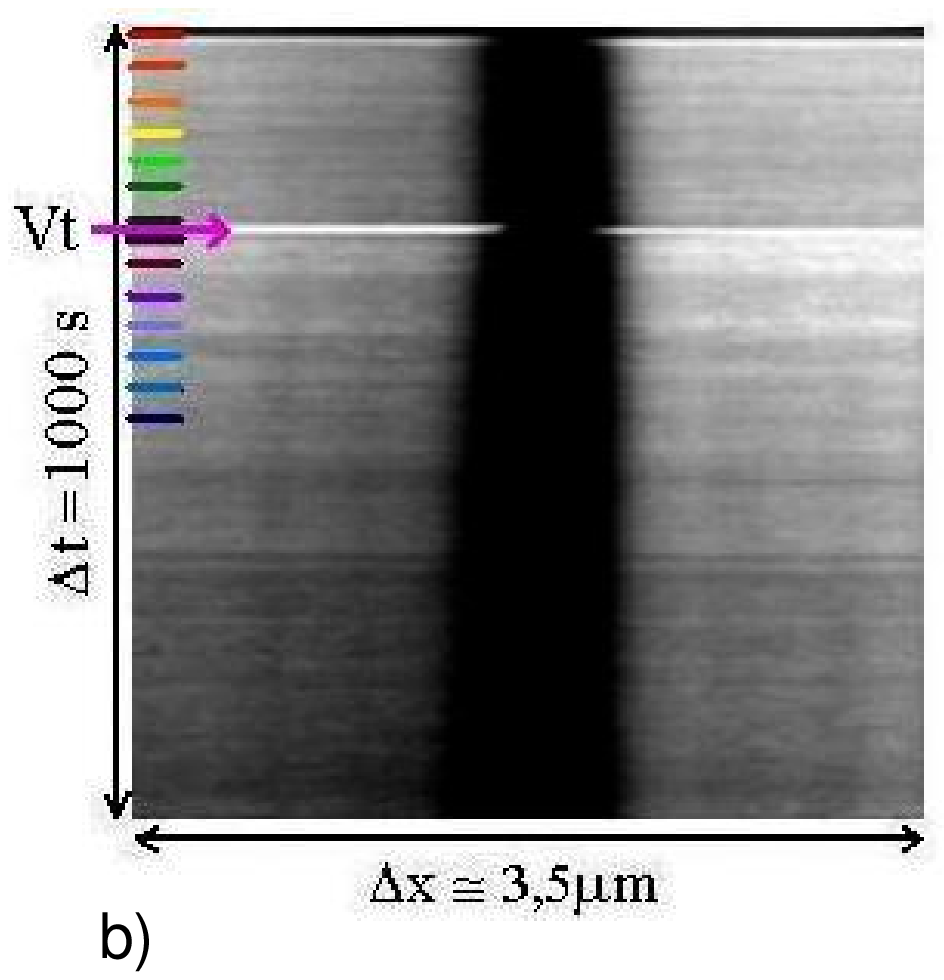}
\includegraphics[width=4cm]{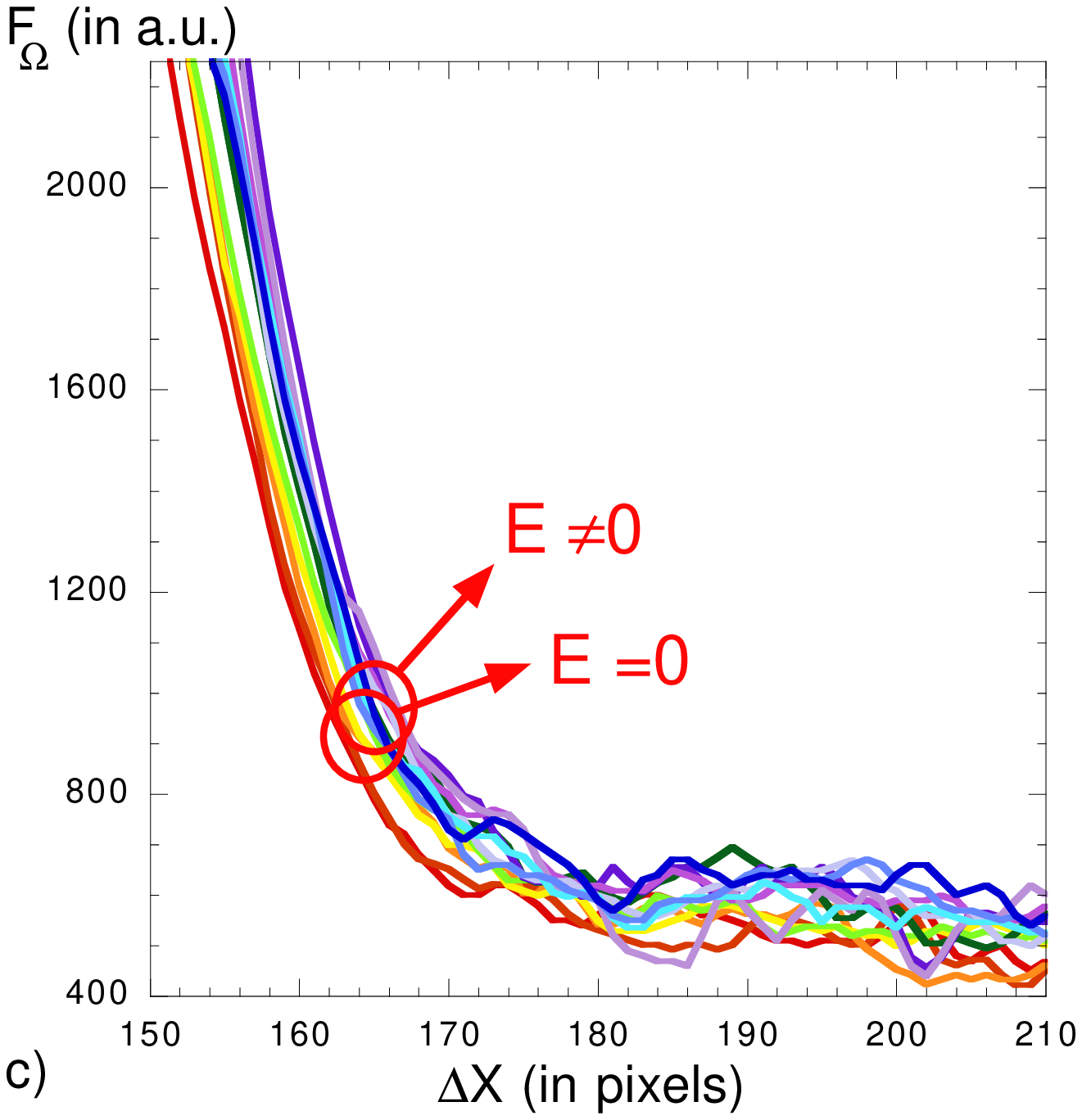}
\caption{\label{f.5} Time-evolution of the electrical signal after a charge deposit ($V_{d}=-100\un{V}$), before and after the switching of transverse electric field. a): detail of the left part of the profile. b): whole profile. c): detail of the right part of the profile.}
\end{center}
\end{figure}
\begin{figure}
    \begin{center}
\includegraphics[width=7cm]{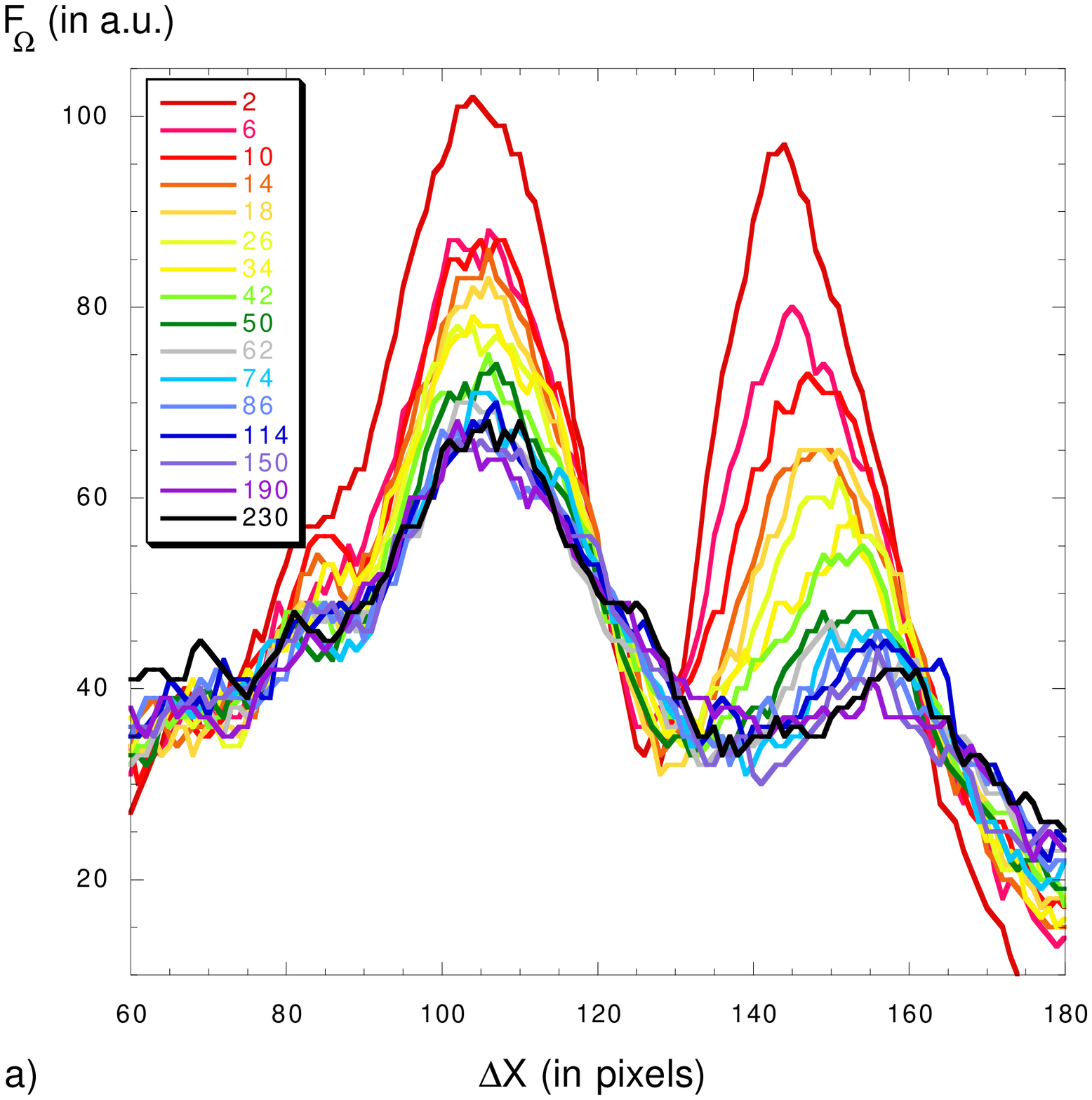}
\includegraphics[width=7cm]{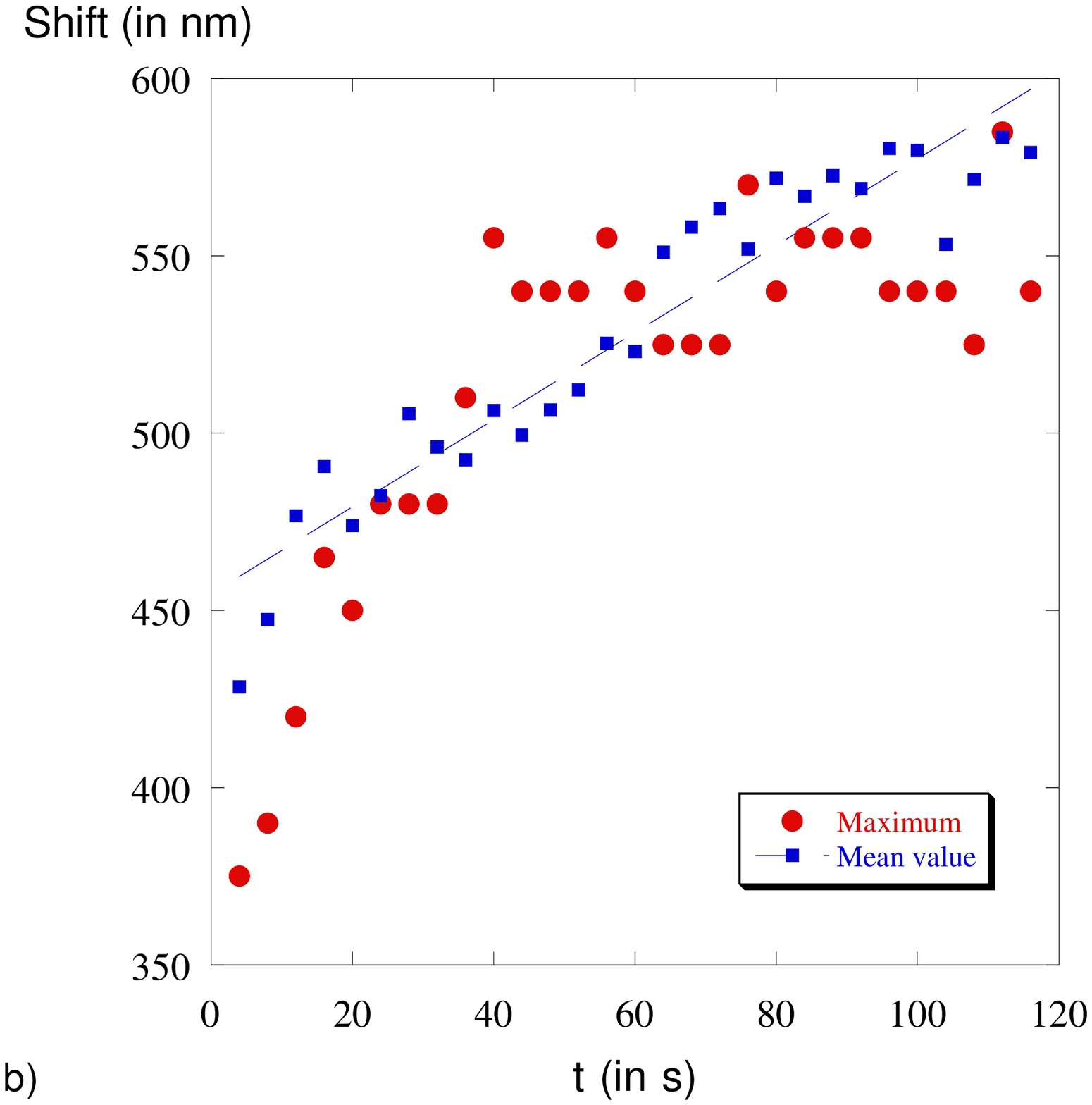}

\caption{a): Time evolution of a "double" charge distribution. After a first charge deposit (on the left), a second one is performed at $V_{d}=-100\un{V}$, $1.5\un{\mu m}$ frome the first one. The behaviors of these two distributions are quite different, the charges belonging to the second one being more mobile. b): Time evolution of the right charge distribution. In red: shift of the maximume. In blue: shift of the first moment of the second charge distribution. The straight line indicates a possible linear dependence.}\label{f.6}
\end{center}
\end{figure}

    The same experiment was repeated in a region between two electrodes on the same sample. An external bias voltage was applied between both electrodes, producing an electric field of about $10^{4}\un{V.m^{-1}}$ near the surface of the layer. Fig.~\ref{f.4}a) shows the evolution of the charge profile with time. Fig.~\ref{f.4}b) shows the evolution of the topography, which was recorded in order to ensure that no drift of the piezoelectric ceramic occurred. One may clearly see the deformation of the charge packet soon after the deposit.
The drift of the charges is characterized by a slight deformation of the profile of the electric signal as well as by a drift of its mean position. A qualitative criterion in order to characterize this drift seems to be the crossing of successive profiles on one side of the charge distribution as may clearly be seen on Fig.~\ref{f.4}. Many similar experiments were performed where the deposited charges exhibited the same qualitative behavior. Fig.~\ref{f.5} shows another way to highlight the drift of the charges in an experiment where the charge relaxation first occurs under no applied electric field before this latter is turned on $200\un{s}$ after the deposit. We focused on the feet on both sides of the charge distribution: As one side of the charge distribution does not move (right side on Fig.~\ref{f.5}), the other moves slightly. This experiment leads us to emphasize the fact that only a small fraction of the deposited charges really move significantly under the transverse electric field. This fraction diminishes as the relaxation across the layer goes on. Thus it is necessary to focus on the feet of the distribution in order to characterize the movement of this small fraction of the deposited charges.\\
This charge displacement can be amplified if the deposit is performed on an initially charged surface. Indeed we noticed that the drift of a charge packet was more important when the surface had previously been charged.
Therefore, in order to maximize the number of charges involved in the transverse displacement of the packet,
an experiment involving two successive charge deposits was performed. A first charge deposit is performed by
applying a deposit voltage $V_{d}= -10 \un{V}$ between the counter-electrode and the tip during
$1500\un{ms}$, in order to deposit a few thousand positive charges (i.e. remove a few thousand electrons) on
(from) the surface. A second charge deposit is performed 100 s after the first one, in the same conditions.
The distance between the two deposits is approximately $1\un{\mu m}$. Fig.~\ref{f.6}a) shows the time evolution of
the electrical signal along a segment crossing the center of the two charge deposits, the origin of time
being taken at the moment of the second deposit. The first deposit is visible on the left of Fig.~\ref{f.6}a), and
one may first notice that its amplitude is the same in the beginning of the scanning as the amplitude of the
second deposit. Yet the behaviors of the two sets of charges are different: A small fraction of the charges
of the first deposit only move under the force exerted by the electric field whereas a bigger fraction of the
charges belonging to the second set move under the same electric field. This observation is made clear from
the fact that the maximum of the first charge distribution does not seem to move, whereas one may measure the
drift of the maximum of the second charge deposit. Thus one may conclude that the charges of the second
deposit are more mobile than the charges belonging to the first deposit. We give an interpretation of this
phenomenon in the following section.
 In order to confirm the nature of the mobile charges and to characterize more precisely their transport from
our measurements, we computed the time-evolution of the displacement of the maxima of the charge packets
together with the first momentum of the single-variable distribution extracted from the measurements of the
profiles. This procedure might seem risky since the total number of charges involved in the transverse
transport may not be considered as constant because of the disappearing of some charges in the
counter-electrode. Yet, in a first approximation, we suppose here that the transverse and in-depth transport
are decoupled, and thus that the fraction of charges involved in the transverse transport is constant at any
time, even if the total number of these charges decreases with time. This hypothesis is reinforced by the
experimental evidence that the free charge distribution previously described does not spread, although the
lateral, or transverse electric field, is by no means homogeneous as one goes from the middle to the borders
of the distribution: the charges seem to drift towards the counter-electrode at the same speed regardless of
the transverse electric field they are submitted to. Fig.~\ref{f.6}b) shows the time evolution of both quantities. In
the canonical case of diffusive transport the evolution of the maximum and of the mean of the packet should
be parallel. Its is not the case here. Fitting linearly the evolution of the first momentum gives an
estimation of the mobility of the charges under the transverse electric field:
$\mu=0.15\un{cm^2.s^{-1}.V^{-1}}$; a value which is less by two orders of magnitude to the one usually
reported for electrons in the conduction band in \chem{SiO_{2}} (e.g. Mott gives $\mu_{e}=20
\un{cm^2.s^{-1}.V^{-1}}$ in \cite{mott1987}), but which is much larger than the value of the mobility
reported for holes in the same material ($\mu_{h}=10^{-6}\un{cm^2.s^{-1}.V^{-1}}$). This result is compatible
with our assumption that the moving charges are electrons hopping from trap states to trap states below the
conduction band.

\section{Conclusion} 
In order to conclude, let us recall the main experimental results presented in this letter : Without an applied transverse electric field, the charges deposited on the silicon oxide surface are attracted by their images in the counter-electrode and move inside the layer. We show in particular that no spreading of the charge packet is observed. during this migration, there is no diffusion along the surface. A very different behavior can be observed when a transverse field is applied between the two embedded electrodes. In this case, deformation of a deposited  charge packet was clearly identified and measured. Its  amplitude depends on the state of charge of the insulating surface : the drift is larger in the case of initially charged surfaces. This results indicate that this lateral displacement is mediated by electric field through the traps located in the oxide surface (over a width about few tens of nm) in the oxide and that the dynamic characteristics of the charge packet is controlled by the trapping-escaping processes of the electrons. 
Moreover, with this experiment, we can answer to the important question of the nature of the transport in such systems. We have restricted ourselves to a qualitative estimation of the transport by noticing the strong spreading and loss of symmetry of the second charge distribution, indicating that the transport is not diffusive if not completely dispersive.  A more precise description of the transport would require the computation of the following momenta of the distribution in order to characterize precisely the spreading, the loss of symmetry and flattening of the charge distribution (For example, in the case of dispersive transport, Scher and Montroll found that the time evolution of the first and second momenta were alike) .  Hence, this observation of the displacement of deposited charges along insulating  surface in the direct space allows to discuss more precisely (without ambiguity) the electron behavior in insulator surfaces containing traps, so we believe this kind of analysis is promising in both fundamental and technological areas.




\acknowledgments  


\end{document}